\documentclass[12pt]{article}
\usepackage{amsmath}
\usepackage{amssymb}
\usepackage{epsfig}
\usepackage{amsthm}

\numberwithin{equation}{section}
\numberwithin{figure}{section}

\def\eq#1{(\ref{eq:#1})}
\def\lineup{\!\!\!\!\!\!\!\!&&}

\newcommand{\Tr}{\mathop{\rm Tr}\nolimits}

\def\d{\partial}

\def\fraction#1#2{ { \textstyle \frac{#1}{#2} }}
\def\half{\fraction{1}{2}}
\def\Box#1{\boxed{\phantom{\Bigg(}#1\ \ \ }}

\def\eps{\epsilon}
\def\beps{\bar{\epsilon}}

\begin{document}

\begin{titlepage}

\begin{center}

\vskip 1.0cm {\large \bf{The Phantom Term in Open String Field Theory}}
\\
\vskip 2.0cm

{\large Theodore Erler$^{(a)}$\footnote{Email: tchovi@gmail.com}}
\vskip .3cm

{\large Carlo Maccaferri$^{(a,b)}$\footnote{Email: maccafer@gmail.com}}

\vskip 1.0cm

$^{(a)}${\it {Institute of Physics of the ASCR, v.v.i.} \\
{Na Slovance 2, 182 21 Prague 8, Czech Republic}}
\vskip .4cm

$^{(b)}${\it INFN, Sezione di Torino\\
Via Pietro Giuria 1, I-10125 Torino, Italy}

\vskip 1.0cm
{\bf Abstract}
\end{center}
We show that given any two classical solutions in open string field theory 
and a singular gauge transformation relating them, it is possible to write the
second solution as a gauge transformation of the first plus a singular,
projector-like state which describes the shift in the open string background 
between the two solutions. This is the ``phantom term.'' We give some
applications in the computation of gauge invariant observables.

\noindent 

\noindent
\medskip

\end{titlepage}

\newpage

\tableofcontents

\baselineskip=18pt

\section{Introduction}

Perhaps the most mysterious aspect of Schnabl's analytic solution for tachyon
condensation \cite{Schnabl} is the so-called {\it phantom term}---a singular 
and formally vanishing term in the solution which appears to be solely 
responsible for the disappearance of the D-brane. Much work has since shed 
light on the term, either specifically in the context of Schnabl's 
solution\cite{Schnabl,Okawa,FKsol}, or in some
generalizations \cite{SSF2,Erler,Russians1,Russians2,simple,exotic}, but so far
there has been limited understanding of why the phantom term should be 
present. 

In this paper we show that the phantom term is a consequence of a particular 
and generic property of string field theory solutions: Given any two solutions
$\Phi_1$ and $\Phi_2$, it is always possible to find a ghost number zero string
field $U$ satisfying
\begin{equation}(Q+\Phi_1)U = U\Phi_2.\end{equation}
This is called a {\it left gauge transformation} from $\Phi_1$ to 
$\Phi_2$ \cite{Integra}. The existence of $U$ implies that $\Phi_2$ can be 
expressed as a gauge transformation of $\Phi_1$ plus a singular projector-like 
state which encapsulates the shift in the open string background between 
$\Phi_1$ and $\Phi_2$. This is the phantom term. The 
phantom term is proportional to a star algebra projector called the 
{\it boundary condition changing projector}, which is conjectured to 
describe a surface of stretched string 
connecting two BCFTs \cite{Integra}. One consequence of this description is 
that phantom terms, in general, do not vanish in the Fock space, as is 
the case for Schnabl's solution.

This paper can be viewed as a companion to reference \cite{Integra}, to which
we refer the reader for more detailed discussion of singular gauge 
transformations and boundary condition changing projectors. Our main goal 
is to show how the phantom term can be used to calculate physical 
observables, even for solutions where the existence of a phantom term was 
not previously suspected. We give three examples: The closed string 
tadpole amplitude \cite{Ellwood} for identity-like marginal deformations 
\cite{rolling}; the energy for Schnabl's solution 
\cite{Schnabl}; and the shift in the closed string tadpole amplitude
between two Schnabl-gauge marginal solutions \cite{RZOK,Schnabl2}. The last 
two computations reproduce results which have been obtained in other 
ways \cite{Schnabl,Kishimoto}, but our approach brings a different perspective 
and some simplifications. This description of the 
phantom term will be useful for the study of future solutions.

\section{The Phantom Term}
\label{sec:phantom}

To start, let's review some concepts and terminology from \cite{Integra}. 
Given a pair of classical solutions $\Phi_1$ and $\Phi_2$, a 
ghost number zero state $U$ satisfying 
\begin{equation}(Q+\Phi_1)U=U\Phi_2\label{eq:left}\end{equation}
is called a {\it left gauge transformation} from $\Phi_1$ to $\Phi_2$. 
If $U$ is invertible, then $\Phi_1$ and $\Phi_2$ are gauge equivalent 
solutions. $U$, however, does not need to be invertible. 
In this case, we say that the left gauge transformation 
is a {\it singular gauge transformation}. 

It is always possible to relate any pair of solutions by a left gauge 
transformation. Given a ghost number $-1$ field $b$, we can construct 
a left gauge transformation from $\Phi_1$ to $\Phi_2$ explicitly with the 
formula: 
\begin{eqnarray}U\lineup = Q b +\Phi_1 b+b\Phi_2\nonumber\\
\lineup=Q_{\Phi_1\Phi_2}b.
\label{eq:exact}\end{eqnarray}
Here, $Q_{\Phi_1\Phi_2}$ is the shifted kinetic operator for a stretched 
string between the solutions $\Phi_1$ and $\Phi_2$. Equation \eq{exact} is not 
necessarily the most general left gauge transformation from $\Phi_1$ to 
$\Phi_2$. This depends on whether $Q_{\Phi_1\Phi_2}$ has cohomology at 
ghost number zero \cite{Integra}.

In the examples we have studied, the left gauge transformation $U$ has an 
important property: If we add a small positive constant to $U$, the resulting 
gauge parameter $U+\eps$ is invertible.\footnote{This is true with the 
appropriate choice of sign for $U$. It would be interesting to better 
understand why this property holds.} This raises a question: If an 
infinitesimal modification of $U$ can make it invertible, why are $\Phi_1$ 
and $\Phi_2$ not gauge equivalent? The answer to this question is contained 
in the following identity:
\begin{equation}\Box{\Phi_2 \,= \, 
\underbrace{\phantom{\Bigg(}\frac{1}{\eps+ U}\Big[\,Q+\Phi_1\,\Big](\eps+ U)
\phantom{\Bigg(}}_{\displaystyle 
\equiv \Phi_1(\eps)}
\, +\, 
\underbrace{\phantom{\Bigg(}
\frac{\eps}{\eps+ U}(\Phi_2 - \Phi_1)\phantom{\Bigg(}
}_{\displaystyle 
\equiv\psi_{12}(\eps)}
,}\label{eq:pgph}\end{equation}
which follows easily from the definition of $U$. The first term 
$\Phi_1(\eps)$ is a gauge transformation of $\Phi_1$, and 
the second term $\psi_{12}(\eps)$ is a remainder. Apparently, if $\Phi_1$ and 
$\Phi_2$ are not gauge equivalent, the remainder must be nontrivial 
in the $\eps\to 0$ limit:
\begin{equation}\lim_{\eps\to0^+}\psi_{12}(\eps)=X^\infty(\Phi_2-\Phi_1).
\label{eq:ph}\end{equation} This is the {\it phantom term}.  The phantom term
is proportional to a star algebra projector,
\begin{equation}X^\infty = \lim_{\eps\to 0^+}\frac{\eps}{\eps+U},
\label{eq:bcc}\end{equation}
called the {\it boundary condition changing projector} \cite{Integra}. 
The boundary condition changing projector is a subtle object, 
and is responsible for some of the ``mystery'' of the phantom term. Based
on formal arguments and examples, it was argued in \cite{Integra} that 
the boundary condition changing projector represents a surface of stretched 
string connecting the BCFTs of $\Phi_2$ and $\Phi_1$. 

The phantom term is useful because it gives an efficient method for computing 
gauge invariant observables from classical solutions. 
In the $\eps\to 0$ limit, the pure-gauge term $\Phi_1(\eps)$ effectively 
``absorbs'' all of the gauge-trivial artifacts of the solution, leaving the 
phantom term to describe the shift in the open string background in a 
transparent manner. In this paper we evaluate the on-shell action and the 
closed string tadpole amplitude \cite{Ellwood}:
\begin{equation}S[\Phi_2] = -\frac{1}{6}\Tr\big[\Phi_2 Q\Phi_2\big],\ \ 
\ \ \ \ \ \ \ 
\Tr_\mathcal{V}[\Phi_2] = 
\mathcal{A}_2(\mathcal{V})-\mathcal{A}_0(\mathcal{V}).
\end{equation}
Here we use the notation,
\begin{eqnarray}
\mathcal{A}_0(\mathcal{V})\lineup
={\mathrm{Disk\ tadpole\ amplitude\ for\ an\ 
on\ shell\ closed\ string\ }\atop 
\mathcal{V}=c\tilde{c}\mathcal{V}^\mathrm{m}\  \mathrm{coupling\ to\ the\ 
reference\ BCFT,}}\nonumber\\
\mathcal{A}_2(\mathcal{V})\lineup 
=\mathrm{Same\ as\ } \mathcal{A}_0(\mathcal{V})\ 
\mathrm{but\ coupling\ to\ the\ BCFT\ of\ } \Phi_2, \nonumber\\
\Tr[\cdot]\lineup = \mathrm{1\mbox{-}string\ vertex\ (the\ Witten\ integral)},
\nonumber\\
\Tr_\mathcal{V}[\cdot]\lineup = \mathrm{1\mbox{-}string\ vertex\ with\ 
midpoint\
insertion\ of\ } \mathcal{V}.\nonumber
\end{eqnarray}
In the subalgebra of wedge states with insertions, the 1-string vertex 
$\Tr[\cdot]$ is equivalent to a correlation function on the cylinder 
\cite{RSZproj} whose circumference is determined by the total wedge angle 
(cf. appendix A of \cite{simple}). The shift in the action and the tadpole 
between the solutions $\Phi_1$ and $\Phi_2$ can be conveniently expressed 
using the phantom term:
\begin{equation}\Box{S[\Phi_2]-S[\Phi_1] = 
\frac{1}{6}\Tr\Big[\psi_{12}(\eps)Q\psi_{12}(\eps)\Big] 
-\frac{1}{3}\Tr\Big[\psi_{12}(\eps)Q\Phi_2\Big]},
\label{eq:action}\end{equation}
and
\begin{equation}\Box{\mathcal{A}_2(\mathcal{V})-\mathcal{A}_1(\mathcal{V})
=\Tr_\mathcal{V}\big[\psi_{12}(\eps)\big]}.\label{eq:ov}\end{equation}
These equations are exact for any $\eps$, though they are most 
useful in the $\eps\to 0$ limit. It would be interesting to see whether the 
phantom term can also be useful for computing the boundary state 
\cite{boundary}. 

Note that the phantom term is a property of a {\it pair} of solutions and 
a singular gauge transformation relating them. In this sense, a solution
by itself does not have a phantom term. That being said, some 
solutions---like Schnabl's solution---seem to be naturally defined as a 
limit of a pure gauge configuration subtracted against a phantom term.
Other solutions, like the ``simple'' tachyon vacuum \cite{simple} and marginal
solutions, can be defined directly without reference to a singular gauge 
transformation or its phantom term. It would be interesting to 
understand what distinguishes these two situations, and why.

\section{Relation to Schnabl's Phantom Term}

The phantom term defined by equations \eq{pgph} and \eq{ph} is 
different from the phantom term as it 
conventionally appears in Schnabl's solution \cite{Schnabl} or some of its 
extensions \cite{SSF2,Erler,exotic}. The standard phantom term can be 
derived from the identity 
\begin{equation}\Phi_2 = \frac{1-X^N}{U}(Q+\Phi_1)U+X^N\Phi_2.
\label{eq:pgph_Sch}\end{equation}
where we define $X$:
\begin{equation}U\equiv 1-X.\end{equation} 
In the $N\to\infty$ limit, the second term in \eq{pgph_Sch} is the phantom 
term:
\begin{equation}\lim_{N\to\infty}X^N\Phi_2 = X^\infty\Phi_2.\end{equation}
This is different from \eq{ph}, though both phantom terms are 
proportional to the boundary condition changing projector. The major 
difference between the identities \eq{pgph} and \eq{pgph_Sch} is 
that $\Phi_1(\eps)$ is {\it exactly} gauge equivalent to $\Phi_1$ for all 
$\eps>0$, whereas the corresponding term in \eq{pgph_Sch},
\begin{equation}\frac{1-X^N}{U}(Q+\Phi_1)U,\end{equation}
is not gauge equivalent to $\Phi_1$, or even a solution, for any finite 
$N$. In this sense \eq{pgph} is a more natural, 
and this is the definition of the phantom term we will use in subsequent 
computations. However, the phantom term can be defined in many ways using 
many different identities similar to \eq{pgph} and \eq{pgph_Sch}, and for 
certain purposes some definitions may prove to be more convenient than others.

To make the connection to earlier work, let us explain how the identity
\eq{pgph_Sch} leads to the standard definition of Schnabl's solution as 
a regularized sum subtracted against a phantom term. We can use Okawa's left 
gauge transformation\footnote{See \cite{Okawa,SSF1} and appendix A of 
\cite{simple} for explanation of the algebraic notation for wedge states 
with insertions which we employ.}
\begin{equation}U=1-\sqrt{\Omega}cB\sqrt{\Omega},\end{equation}
to map from the perturbative vacuum $\Phi_1=0$ to Schnabl's solution
\begin{equation}\Phi_2=\Psi = \sqrt{\Omega}c\frac{KB}{1-\Omega}c\sqrt{\Omega}.
\end{equation}
Substituting these choices into \eq{pgph_Sch} gives the expression 
\begin{equation}\Psi = -\sum_{n=0}^{N-1}\psi_n' 
+\sqrt{\Omega}c\Omega^N\frac{KB}{1-\Omega}c\sqrt{\Omega},\label{eq:sch1}
\end{equation}
where \cite{Schnabl}
\begin{equation}\psi_n' \equiv \frac{d}{dn}\psi_n,\ \ \ 
\psi_n \equiv \sqrt{\Omega}cB\Omega^nc\sqrt{\Omega}.\label{eq:psin}\end{equation}
To simplify further, expand  
\begin{equation}\frac{K}{1-\Omega} = \sum_{n=0}^\infty \frac{B_n}{n!}(-K)^n,
\label{eq:Bn}\end{equation}
inside the second term of \eq{sch1}, where $B_n$ are the Bernoulli numbers. 
The expansion in powers of $K$ is equivalent to the $\mathcal{L}^-$ level 
expansion 
\cite{exotic,RZ},\footnote{$\mathcal{L}^-$ is the BPZ odd component of 
Schnabl's $\mathcal{L}_0$, 
the zero mode of the energy momentum tensor in the sliver coordinate 
frame \cite{Schnabl,RZ}. It is
a derivation and a reparameterization generator, and computes 
scaling dimension of operator insertions in correlation functions on the 
cylinder. See \cite{exotic} for discussion of the $\mathcal{L}^-$ level 
expansion.} which will play an important role in simplifying
correlators involving the phantom term. The upshot in the current context is 
that the higher powers of $K$ in \eq{Bn} can usually be ignored in the 
$N\to\infty$ limit \cite{Erler}, so we can effectively replace the sum by 
its first term
\begin{equation}\frac{K}{1-\Omega} \to 1.\label{eq:lead}\end{equation}
Then the $N\to\infty$ limit of \eq{pgph_Sch} reproduces the usual
expression for Schnabl's solution
\begin{equation}\Psi = \lim_{N\to\infty}\left[-\sum_{n=0}^N\psi_n'+\psi_N
\right],\label{eq:Sch}\end{equation}
where $\psi_N$ is the phantom term.

\section{Example 1: Identity-like Marginals}
\label{sec:warm}

We start with a simple example: computing the shift in the closed string 
tadpole amplitude between the identity-like solution for the tachyon vacuum 
\cite{simple,id1,id2},
\begin{equation}\Phi_1 = c(1-K),\label{eq:idtv}\end{equation}
and the identity-like solution for a regular marginal 
deformation \cite{rolling},
\begin{equation}\Phi_2=cV,\label{eq:id}\end{equation}
where $V$ is a weight $1$ matter primary with regular OPE with itself. 
Both these solutions are singular. For example, we cannot evaluate 
the tadpole directly because
\begin{equation}\Tr_\mathcal{V}[cV]\end{equation}
requires computing a correlator on a surface with vanishing area. 
However, with the phantom term, we can circumvent this problem with a few 
formal (but natural) assumptions. 

We can relate the above solutions with a left gauge transformation
\begin{eqnarray}U \lineup = Q_{\Phi_1\Phi_2}B\nonumber\\
\lineup = 1+Bc(K+V-1).\end{eqnarray}
The shift in the open string background is described by the phantom 
term:\footnote{In the following examples the 
dependence on $\eps$ simplifies if we make a reparameterization 
$\eps\to\eps/\beps$ relative to \eq{pgph}, where by definition 
$\beps\equiv 1-\eps$.} 
\begin{eqnarray}\psi_{\Phi_1\Phi_2}(\eps) \lineup = 
\frac{\eps}{\eps+\beps U} (\Phi_2-\Phi_1)\ \ \ \ \ \ \ \ \ \ \ \ \ \ 
(\beps\equiv 1-\eps)\nonumber\\
\lineup 
= \left(-\eps c+\frac{\eps\beps}{\eps +\beps(K+V)}Bc\d c \right)(1-K-V)
\nonumber\\ \lineup
= \left(-\eps c+\beps \int_0^\infty dt\, e^{- t}\ 
\Omega_V^{\ \beps t/\eps}Bc\d c \right)(1-K-V).
\label{eq:Vph}\end{eqnarray} 
where in the third line we defined the states
\begin{equation}\Omega_V^{\ t} \equiv e^{-t(K+V)}.\end{equation}
These are wedge states whose open string boundary conditions have been 
deformed by the marginal current $V$ \cite{bcc}. In the $\eps\to 0$ limit
the phantom term is
\begin{equation}\lim_{\eps\to 0}\psi_{\Phi_1\Phi_2}(\eps) = 
\Omega_V^\infty Bc\d c (1-K-V),\end{equation}
Note that the phantom term corresponds to a nondegenerate surface with the 
boundary conditions of the marginally deformed BCFT, and so will
naturally reproduce the expected coupling to closed strings. This is in spite
of the fact that both solutions we started with were identity-like. Also 
note that the phantom term vanishes in the Fock space (since $B$ kills the 
sliver state), but still it is nontrivial. 

Now we can use the phantom term to compute the shift in the closed string 
tadpole amplitude:
\begin{equation}\Tr_\mathcal{V}[\psi_{\Phi_1\Phi_2}(\eps)].\label{eq:Vov}
\end{equation}
Since the amplitude vanishes around the tachyon vacuum, only the 
marginally deformed D-brane should contribute. Plugging \eq{Vph} in, we find
\begin{equation}\Tr_\mathcal{V}[\psi_{\Phi_1\Phi_2}(\eps)]
=\Tr_\mathcal{V}\left[
\,\eps cB(K+V-1)c\,+\,\frac{\eps}{\eps+\beps (K+V)}Bc\d c\right].
\label{eq:1st}\end{equation}
The first term in the trace formally vanishes.\footnote{This first term 
in \eq{1st} formally vanishes because it is the trace of a state which 
has negative scaling dimension plus a state which is BRST exact with respect 
to the BRST operator of the marginally deformed BCFT. However, rigorously 
speaking the trace is undefined, since computing it requires evaluating a 
correlator on a surface with vanishing area. This 
is a remnant of the fact that our marginal solution and tachyon vacuum are 
too ``identity-like.'' However, since the offending term is proportional to 
$\eps$, and \eq{Vov} is formally independent of $\eps$, we will set $\eps=0$ 
and ignore this term.} Then
\begin{eqnarray}\Tr_\mathcal{V}[\psi_{\Phi_1\Phi_2}(\eps)]
\lineup =\Tr_\mathcal{V}\left[\frac{\eps}{\eps+\beps (K+V)}Bc\d c\right]
\nonumber\\
\lineup 
=\int_0^\infty dt\, e^{- t}\,
\Tr_\mathcal{V}[\Omega_V^{\ \beps t/\eps }Bc\d c].
\end{eqnarray}
With the reparameterization we can scale the deformed wedge state inside 
the trace to unit width:
\begin{equation}\Tr_\mathcal{V}[\Omega_V^{\beps t/\eps}Bc\d c] = 
\Tr_\mathcal{V}\left[\left(\frac{\beps t}{\eps}\right)^{\frac{1}{2}\mathcal{L}^-}\Big(\Omega_V
B c\d c\Big)\right]= \Tr_\mathcal{V}[\Omega_V Bc\d c].
\end{equation}
Integrating over $t$ gives 
\begin{equation}\Tr_\mathcal{V}[\psi_{\Phi_1\Phi_2}(\eps)]
=\Tr_\mathcal{V}\left[\Omega_V Bc\d c\right].\label{eq:Vm1}\end{equation}
Note that this is manifestly independent of $\eps$. In the general situation, 
explicitly proving $\eps$-independence requires much more work than is needed 
to compute the result, and it is easier to assume gauge invariance and 
take the $\eps\to 0$ limit. At any rate, further simplifying \eq{Vm1},
we can replace the ghost factor $Bc\d c$ in the trace with 
$-c$,\footnote{This follows from the 
fact that the derivation 
$\mathcal{B}^-$ annihilates the 1-string vertex\cite{simple}, and
\begin{equation}-\half\mathcal{B}^-\big(\Omega_V c\d c\big) 
= \Omega_V Bc\d c 
+\Omega_V c.\label{eq:Bghost}\end{equation}}. Mapping from the cylinder 
to the unit disk gives
\begin{equation}\Tr_\mathcal{V}[\psi_{\Phi_1\Phi_2}(\eps)] = -\frac{1}{2\pi i}
\left\langle\exp\left[\int_0^{2\pi}d\theta\, V(e^{i\theta})\right]
\mathcal{V}(0)
c(1)\right\rangle_{\mathrm{disk}}.\end{equation}
This is exactly the closed string tadpole amplitude for the marginally deformed
D-brane, as defined in the conventions of \cite{Ellwood}. 

\section{Example 2: Energy for Schnabl's Solution}
\label{sec:Schnabl}

In this section we compute the energy for Schnabl's solution,
\begin{equation}\Psi = \sqrt{\Omega}c\frac{KB}{1-\Omega}c\sqrt{\Omega}.
\end{equation}
The original computation of the energy, based on the expression \eq{Sch},
was given in \cite{Schnabl} (see also \cite{Okawa,FKsol}). 
Our computation will 
be quite different since we define the phantom term in a different way. 

We take the reference solution to be the perturbative vacuum, and map to
Schnabl's solution using Okawa's left gauge transformation
\begin{eqnarray}U\lineup = 1-\sqrt{\Omega}cB\sqrt{\Omega}\nonumber\\
\lineup = Q_{0\Psi}\left(B\frac{1-\Omega}{K}\right).
\end{eqnarray}
The regularized phantom term is
\begin{equation}\psi_{0\Psi}(\eps) = \frac{\eps}{\eps+\beps U}\Psi
=\sqrt{\Omega}cB\frac{\eps}{1-\beps\,\Omega}
\frac{K}{1-\Omega}c\sqrt{\Omega}.\label{eq:ph_Sch}\end{equation}
In the $\eps\to 0$ limit the ratio $\frac{\eps}{1-\beps \Omega}$ approaches 
the sliver state (see later), so we can replace the factor $\frac{K}{1-\Omega}$
with its leading term in the $\mathcal{L}^-$ level expansion. Then \eq{pgph}
gives a regularized definition of Schnabl's solution:
\begin{equation}\Psi = \lim_{\eps\to 0^+}\left[\Psi_{\beps} +
\sqrt{\Omega}cB\frac{\eps}{1-\beps\,\Omega}c\sqrt{\Omega}\right].
\label{eq:Sch2}\end{equation} 
Note that $\Psi_{\beps}$ here is precisely the pure gauge solution discovered
by Schnabl \cite{Schnabl}: 
\begin{equation}\Psi_{\beps} = \beps 
\sqrt{\Omega}c\frac{KB}{1-\beps\,\Omega}c\sqrt{\Omega}.\end{equation}
Using \eq{psin} we can express this regularization in the form
\begin{equation}\Psi = \lim_{\eps\to 0^+}\sum_{n=0}^\infty
\beps^{\,n}\Big[-\psi_n'+\eps\,\psi_n\Big].\end{equation}
Clearly this is different from the standard definition of Schnabl's 
solution, \eq{Sch}.

To calculate the action we use \eq{action}:
\begin{equation}S = \frac{1}{6}\Tr\Big[\psi_{0\Psi}(\eps)
Q\psi_{0\Psi}(\eps)\Big]-\frac{1}{3}\Tr\Big[\psi_{0\Psi}(\eps)Q\Psi\Big].
\end{equation}
A quick calculation shows that the second term can be ignored in the 
$\eps\to 0$ limit, essentially because the phantom term vanishes when 
contracted with well-behaved states. Therefore
\begin{equation}S = \frac{1}{6}\lim_{\eps\to 0^+}
\Tr\Big[\psi_{0\Psi}(\eps)\,Q\psi_{0\Psi}(\eps)\Big].\end{equation}
Substituting the phantom term \eq{ph_Sch} gives an expression of the form
\begin{equation}\Tr\Big[\psi_{0\Psi}(\eps)\,Q\psi_{0\Psi}(\eps)\Big]=
-\Tr\left[C_1\frac{\eps}{1-\beps\,\Omega}C_2\frac{\eps}{1-\beps\,\Omega}\right]
+\Tr\left[C_3\frac{\eps}{1-\beps\,\Omega}C_4\frac{\eps}{1-\beps\,\Omega}
\right],\label{eq:Sch_step1}\end{equation}
where 
\begin{eqnarray}
C_1\ \lineup =\ [c,\Omega]\frac{KB}{1-\Omega},\\
C_2\ \lineup =\ c\Omega c\d c\frac{K}{1-\Omega},\\
C_3\ \lineup =\ cK[c,\Omega]\frac{KB}{1-\Omega},\\
C_4\ \lineup =\ c\Omega c\frac{K}{1-\Omega}.
\end{eqnarray}
To understand what happens in the $\eps\to 0$ limit, note that the 
factor $\frac{\eps}{1-\beps \Omega}$ inside the trace \eq{Sch_step1} 
approaches the sliver state:
\begin{equation}\lim_{\eps\to 0^+}\frac{\eps}{1-\beps\Omega} = \Omega^\infty.
\end{equation} 
To prove this, expand the geometric series 
\begin{equation}\frac{\eps}{1-\beps\Omega}=
\eps\sum_{n=0}^\infty \beps^n \Omega^n.\end{equation}
and expand the wedge state in the summand around $n=\infty$:
\begin{equation}\Omega^n = \Omega^\infty +\frac{1}{n+1}\Omega_{(1)}
+\frac{1}{(n+1)^2}\Omega_{(2)}+\ .\ .\ .\ ,
\label{eq:slivexp}\end{equation}
where $\Omega_{(1)},\Omega_{(2)},...$ are the coefficients of the corrections
in inverse powers of $n+1$ (actually $\Omega_{(2)}$ is the first nonzero 
correction in the Fock space). Plugging \eq{slivexp} into the geometric 
series and performing the sums gives
\begin{equation}\frac{\eps}{1-\beps\Omega}= \Omega^\infty 
+\frac{\eps\ln\eps}{\beps}\Omega_{(1)}
+ \frac{\eps\mathrm{Li}_2\beps}{\beps}\Omega_{(2)}+\ .\ .\ .\ .
\label{eq:slivexp2}
\end{equation}
Only the sliver state survives the $\eps\to 0$ limit. This means that for 
small $\eps$ equation \eq{Sch_step1} is dominated by correlation functions 
on the cylinders with very large circumference. In this limit, it is useful to
expand the fields $C_1,...C_4$ into a sum of states with definite scaling 
dimension (the $\mathcal{L}_0$ level expansion). To leading order this 
expansion gives 
\begin{eqnarray}
C_1\lineup = 
\sqrt{\Omega}(\d c B)\sqrt{\Omega} + .\ .\ .\ ,
\ \ \ \ \ \ \ \ \ \ \ \ \ \ \ \ \  (\mathcal{L}_0 = 1), \\
C_2\lineup = -\frac{1}{2}\sqrt{\Omega}(c\d c\d^2 c)\sqrt{\Omega}
+ .\ .\ .\ ,
\ \ \ \ \ \ \ \ \ \ \!(\mathcal{L}_0 = 0), \\
C_3\lineup = \sqrt{\Omega}(Kc\d cB)\sqrt{\Omega}
+ .\ .\ .\ ,
\ \ \ \ \ \ \ \ \ \ \ \ \ \ \!\!(\mathcal{L}_0 = 1), \\
C_4\lineup = -\sqrt{\Omega}(c\d c)\sqrt{\Omega} + .\ .\ .\ ,
\ \ \ \ \ \ \ \ \ \ \ \ \ \ \ \ \  \!\!(\mathcal{L}_0 = -1).
\end{eqnarray}
Now consider the following: If a cylinder of circumference $L$ has 
insertions of total scaling dimension $h$ separated parametrically with $L$, 
rescaling the cylinder down to unit circumference produces an overall factor 
of $L^{-h}$, which vanishes in the large circumference limit if $h$ is 
positive. Since the sum of the lowest scaling dimensions of $C_1$ and $C_2$ is 
positive, the corresponding term in \eq{Sch_step1} must vanish.
The sum of the lowest scaling dimensions of $C_3$ and $C_4$ is zero, so the 
corresponding term in \eq{Sch_step1} is nonzero and receives contribution only
from the leading $\mathcal{L}_0$ level of $C_3$ and $C_4$. Therefore the 
action simplifies to 
\begin{equation}S = -\frac{1}{6}\lim_{\eps\to 0}
\Tr\left[Kc\d c  B\frac{\eps\, \Omega}{1-\beps\, \Omega}c\d c
\frac{\eps\,\Omega}{1-\beps\,\Omega}\right].\end{equation}
Expanding the geometric series gives
\begin{equation}S = -\frac{1}{6}\lim_{\eps\to 0}
\eps^2 \sum_{L=2}^\infty \beps^L\sum_{k=1}^{L-1}
\Tr\left[Kc\d c  B\Omega^{L-k}c\d c
\Omega^k\right].\end{equation}
Scaling the total wedge angle inside the trace to unity,
\begin{equation}S = -\frac{1}{6}\lim_{\eps\to 0}
\eps^2 \sum_{L=2}^\infty L\beps^L\left(\frac{1}{L}\sum_{k=1}^{L-1}
\Tr\big[Kc\d c  B\Omega^{1-\frac{k}{L}}c\d c
\Omega^{\frac{k}{L}}\big]\right).\end{equation}
Expanding the factor in parentheses around $L=\infty$, the sum turns into an 
integral: 
\begin{equation}\frac{1}{L}\sum_{k=1}^{L-1}
\Tr\big[Kc\d c  B\Omega^{1-\frac{k}{L}}c\d c
\Omega^{\frac{k}{L}}\big]=\int_0^1 dx\, \Tr\big[Kc\d c  B\Omega^{1-x}c\d c
\Omega^x\big]+\mathcal{O}\left(\frac{1}{L}\right)+\ .\ .\ .\ .
\end{equation}
The order $1/L$ terms and higher do not contribute in the $\eps\to 0$ limit,
as explained in \eq{slivexp2}. Therefore
\begin{eqnarray}S \lineup = -\frac{1}{6}
\left(\lim_{\eps\to 0}\eps^2\sum_{L=2}^\infty L\beps^L\right)
\int_0^1 dx\, \Tr\big[Kc\d c  B\Omega^{1-x}c\d c
\Omega^x\big]\nonumber\\
\lineup =-\frac{1}{6}\int_0^1 dx\,
\Tr\Big[\Omega^{1-x}Bc\d c\, Q\big(\Omega^x Bc \d c\big)\Big].
\end{eqnarray}
A moment's inspection reveals that this integral is precisely the action 
evaluated on the ``simple'' solution for the tachyon vacuum \cite{simple},
expressed in the form \cite{BMT}
\begin{equation}\Psi_\mathrm{simp} = c - \frac{B}{1+K}c\d c.\end{equation}
Therefore
\begin{equation}S = -\frac{1}{6}\Tr[\Psi_\mathrm{simp}Q\Psi_\mathrm{simp}] = 
\frac{1}{2\pi^2},\end{equation}
in agreement with Sen's conjecture.

\section{Example 3: Tadpole Shift Between Two Marginals}
\label{sec:Shift}

In this section we use the phantom term to compute the shift in the closed 
string tadpole amplitude between two Schnabl-gauge marginal solutions 
\cite{RZOK,Schnabl2}:
\begin{eqnarray}
\Phi_1 \lineup = \sqrt{\Omega}cV_1\frac{B}{1+\frac{1-\Omega}{K}V_1}c
\sqrt{\Omega},\nonumber\\
\Phi_2 \lineup = \sqrt{\Omega}cV_2\frac{B}{1+\frac{1-\Omega}{K}V_2}c
\sqrt{\Omega},
\end{eqnarray}
where $V_1$ and $V_2$ are weight 1 matter primaries with regular OPEs with
themselves (but not necessarily with each other).\footnote{For example, we 
could choose $V_1= e^{\frac{1}{\sqrt{\alpha'}}X^0}$ to be the rolling tachyon 
deformation, and $V_2= e^{-\frac{1}{\sqrt{\alpha'}}X^0}$ to be the ``reverse''
rolling tachyon; or we could choose $V_1$ and $V_2$ to be Wilson line 
deformations along two independent light-like directions. In both these 
examples the $V_1$-$V_2$ OPE is singular.} 
Our main interest in this example is to understand how the 
boundary condition changing projector works when connecting two distinct and
nontrivial BCFTs; In this case the projector has a rather 
nontrivial structure and possible singularities from the collision of matter 
operators at the midpoint \cite{Integra}. This is the first example 
of a phantom term which does not vanish in the Fock space (at least in the case
where the $V_1$-$V_2$ OPE is regular). This example also gives an independent 
derivation of the tadpole amplitude for Schnabl-gauge marginals, which 
previously proved difficult to compute \cite{Kishimoto}. 
Another computation of the tadpole for the closely related solutions of 
Kiermaier, Okawa, and Soler \cite{KOS} appears in \cite{KOS_super}.

We will map between the marginal solutions $\Phi_1$ and $\Phi_2$ using 
the left gauge transformation
\begin{eqnarray}U\lineup =Q_{\Phi_1\Phi_2}\left(B\frac{1-\Omega}{K}\right) 
\nonumber\\
\lineup = 1-\sqrt{\Omega}cB\frac{1}{1+V_1\frac{1-\Omega}{K}}\sqrt{\Omega}
-\sqrt{\Omega}\frac{1}{1+\frac{1-\Omega}{K}V_2}Bc\sqrt{\Omega}.
\end{eqnarray}
This choice of $U$ is natural to the structure of the marginal solutions, 
since it factorizes into a product of left gauge transformations through 
the Schnabl-gauge tachyon vacuum \cite{Integra}. 
The regularized boundary condition changing projector is
\begin{eqnarray}\frac{\eps}{\eps+\beps U}\lineup = \eps 
+ \underbrace{\phantom{\Bigg(}\beps\sqrt{\Omega}cB
\frac{\eps}{1+V_1\frac{1-\Omega}{K}-\beps\Omega}
\sqrt{\Omega}\phantom{\Bigg(}}_{\displaystyle\equiv P_1}
+ \underbrace{\phantom{\Bigg(}\beps\sqrt{\Omega}
\frac{\eps}{1+\frac{1-\Omega}{K}V_2-\beps\Omega}
Bc\sqrt{\Omega}\phantom{\Bigg(}}_{\displaystyle\equiv P_2}
\nonumber\\
\lineup \ \ \ \ \ \ \ \ 
+ \underbrace{\phantom{\Bigg(}\beps^2\eps\sqrt{\Omega}
\frac{1}{1+\frac{1-\Omega}{K}V_2-\beps\Omega}B[c,\Omega]
\frac{1}{1+V_1\frac{1-\Omega}{K}-\beps\Omega}
\sqrt{\Omega}\phantom{\Bigg(}}_{\displaystyle\equiv P_{12}}.
\end{eqnarray}
The first two terms, $P_1$ and $P_2$, are regularized boundary condition 
changing projectors from $\Phi_1$ to the tachyon vacuum (Schnabl's solution), 
and from the tachyon vacuum to $\Phi_2$, respectively. These terms vanish in 
the Fock space in the $\eps\to 0$ limit. The third term 
$P_{12}$ is the nontrivial one: In the $\eps\to 0$ limit it approaches the 
sliver state in the Fock space, with the boundary conditions of $\Phi_2$ on 
its left half and the boundary conditions of $\Phi_1$ on its right half;
it represents the open string connecting the 
BCFTs of $\Phi_2$ and $\Phi_1$ \cite{Integra}. If $V_1$ and $V_2$ have 
regular OPE, $P_{12}$ is a nonvanishing projector in the Fock space. If the 
OPE is singular, $P_{12}$ may be vanishing or divergent 
because of an implicit singular conformal transformation of the boundary 
condition changing operator between the BCFTs of $\Phi_2$ and $\Phi_1$ at 
the midpoint. Part of our goal is to see how this singularity is resolved 
when we compute the overlap. The phantom term is
\begin{equation}\psi_{12}(\eps) = (\eps +P_1+P_2+P_{12})(\Phi_2-\Phi_1).
\end{equation}
First let us consider the contribution to the tadpole from $P_{12}$:
\begin{equation}\Tr_\mathcal{V}[P_{12}(\Phi_2-\Phi_1)].\end{equation}
Plugging everything in gives 
\begin{eqnarray}
\lineup\!\!\!\!\!\!\!\! \Tr_\mathcal{V}[P_{12}(\Phi_2-\Phi_1)]\nonumber\\
\lineup\!\!\!\!\!\!\!\! = \Tr_\mathcal{V}
\left[\beps^2\eps
\frac{1}{1+\frac{1-\Omega}{K}V_2-\beps\Omega}B[c,\Omega]
\frac{1}{1+V_1\frac{1-\Omega}{K}-\beps\Omega}
\Omega\left(V_2\frac{1}{1+\frac{1-\Omega}{K}V_2} - 
V_1\frac{1}{1+\frac{1-\Omega}{K}V_1}\right)c\Omega\right]\nonumber\\
\lineup\!\!\!\!\!\!\!\! = \beps\eps\Tr_\mathcal{V}
\Bigg[Bc\Omega c
\frac{1}{1+V_1\frac{1-\Omega}{K}-\beps\Omega}
\underbrace{\beps\Omega\left(V_2\frac{1}{1+\frac{1-\Omega}{K}V_2} - 
V_1\frac{1}{1+\frac{1-\Omega}{K}V_1}\right)\Omega}_{\mathrm{next\ step}}
\frac{1}{1+\frac{1-\Omega}{K}V_2-\beps\Omega}\Bigg].\nonumber\\
\label{eq:nxt}\end{eqnarray}
In the second step we inserted a trivial factor of $cB$ next to the commutator
$[c,\Omega]$, which allows us to remove the $c$ ghost from the difference 
between the solutions. Now let's look at the factor above the braces. 
Re-express it with a few manipulations
\begin{eqnarray}\lineup\!\!\!\!\!\!\!\!
\beps\Omega\left(V_2\frac{1}{1+\frac{1-\Omega}{K}V_2} - 
V_1\frac{1}{1+\frac{1-\Omega}{K}V_1}\right)\Omega \nonumber\\
\lineup\!\!\!\!\!\!\!\!= 
-\beps \Omega\left(\frac{1}{1+V_1\frac{1-\Omega}{K}}V_1\frac{1-\Omega}{K}-1
\right)\frac{K\Omega}{1-\Omega}+\beps\frac{K\Omega}{1-\Omega}\left(
\frac{1-\Omega}{K}V_2\frac{1}{1-\frac{1-\Omega}{K}V_2}-1\right)\Omega
\nonumber\\
\lineup\!\!\!\!\!\!\!\!= 
\beps \Omega\frac{1}{1+V_1\frac{1-\Omega}{K}}
\frac{K\Omega}{1-\Omega}-\beps\frac{K\Omega}{1-\Omega}
\frac{1}{1-\frac{1-\Omega}{K}V_2}\Omega\nonumber\\
\lineup\!\!\!\!\!\!\!\!= -\left(1- \beps\Omega\frac{1}{1+V_1\frac{1-\Omega}{K}}
\right)\frac{K\Omega}{1-\Omega}+\frac{K\Omega}{1-\Omega}\left(1-\beps
\frac{1}{1-\frac{1-\Omega}{K}V_2}\Omega\right).\label{eq:shift1}
\end{eqnarray}
Express the factors on either side of the underbrace in equation \eq{nxt} 
in the form:
\begin{eqnarray}
\frac{1}{1+V_1\frac{1-\Omega}{K}-\beps\Omega}\lineup = 
\frac{1}{1+V_1\frac{1-\Omega}{K}}
\left(\frac{1}{1-\beps\Omega\frac{1}{1+V_1\frac{1-\Omega}{K}}}\right)
\nonumber\\
\frac{1}{1+\frac{1-\Omega}{K}V_2-\beps\Omega}\lineup = 
\left(\frac{1}{1-\beps\frac{1}{1+\frac{1-\Omega}{K}V_2}\Omega}\right)
\frac{1}{1+\frac{1-\Omega}{K}V_2}.
\end{eqnarray}
Plugging everything into \eq{nxt}, the factors in parentheses above cancel 
against the factors in parentheses in \eq{shift1}. Thus the contribution to 
the tadpole from $P_{12}$ simplifies to 
\begin{eqnarray}
\Tr_\mathcal{V}[P_{12}(\Phi_2-\Phi_1)] \lineup = 
\beps \Tr_\mathcal{V}\bigg[
\underbrace{\frac{\eps}{1+V_1\frac{1-\Omega}{K}-\beps\Omega}
}_{\displaystyle\to\ {\textstyle \frac{K}{1-\Omega}}\Omega_{V_1}^\infty}
\frac{K\Omega}{1-\Omega}\frac{1}{1+\frac{1-\Omega}{K}V_2}B c\Omega c\bigg]
\nonumber\\
\lineup\ \ 
-\beps\Tr_\mathcal{V}\bigg[Bc\Omega c \frac{1}{1+V_1\frac{1-\Omega}{K}}
\frac{K\Omega}{1-\Omega}
\underbrace{\frac{\eps}{1+\frac{1-\Omega}{K}V_2-\beps\Omega}
}_{\displaystyle\to\ \Omega_{V_2}^\infty{\textstyle\frac{K}{1-\Omega}}}
\bigg].\label{eq:P12}
\end{eqnarray}
In the $\eps\to 0$ limit, we claim that the factors above the braces approach 
the sliver state with boundary conditions deformed by the corresponding 
marginal current, multiplied by the factor $\frac{K}{1-\Omega}$. 
If $V_1$ and $V_2$ have regular OPE, we can expand the 
factors outside the braces in the $\mathcal{L}^-$ level expansion and pick
off the leading term in the $\eps\to0$ limit. This gives precisely the 
difference in the closed string tadpole amplitude between the two solutions. 
Unfortunately this argument does not work when $V_1$ and $V_2$ have 
singular OPE, since contractions between $V_1$ and $V_2$ produce 
operators of lower conformal dimension which make additional contributions. 
It is not an easy task to see what happens in this case in the 
$\eps\to 0$ limit, but there is no reason to believe that \eq{P12} 
should calculate the shift in the tadpole amplitude. 
This is a remnant of the midpoint singularity of the boundary condition 
changing projector when the boundary condition changing operator between the 
BCFTs of $\Phi_2$ and $\Phi_1$ has nonzero conformal weight. 
To fix this problem we need to account for the ``tachyon vacuum'' contributions
to the phantom term. Let us focus on the contribution from $P_1$:
\begin{eqnarray}\Tr_\mathcal{V}[P_1(\Phi_2-\Phi_1)] 
\lineup = \beps\Tr_\mathcal{V}\left[
\frac{\eps}{1+V_1\frac{1-\Omega}{K}-\beps\Omega}\Omega 
V_2\frac{1}{1+\frac{1-\Omega}{K}V_2}Bc\Omega c\right]
-\Tr_\mathcal{V}[P_1\Phi_1]
\nonumber\\
\lineup = \beps\Tr_\mathcal{V}\left[
\frac{\eps}{1+V_1\frac{1-\Omega}{K}-\beps\Omega}\frac{K\Omega}{1-\Omega} 
\left(-\frac{1}{1+\frac{1-\Omega}{K}V_2}+1\right)Bc\Omega c\right]
-\Tr_\mathcal{V}[P_1\Phi_1].\nonumber\\
\end{eqnarray}
Note that this precisely cancels the problematic contractions between 
$V_1$ and $V_2$ in the first term in \eq{P12}. A similar thing happens for 
the second term when we consider the contribution of $P_2$. Therefore, 
in total the overlap is
\begin{eqnarray}\Tr_\mathcal{V}[\psi_{12}(\eps)] \lineup = 
\eps \Tr_\mathcal{V}[\Phi_2-\Phi_1] -\Tr_\mathcal{V}[P_1\Phi_1]
+\Tr_\mathcal{V}[P_2\Phi_2] \nonumber\\
\lineup \ \ +
\beps \Tr_\mathcal{V}\left[
\frac{\eps}{1+V_1\frac{1-\Omega}{K}-\beps\Omega}\frac{K\Omega}{1-\Omega}
B c\Omega c\right]
-\beps\Tr_\mathcal{V}\left[Bc\Omega c\frac{K\Omega}{1-\Omega}
\frac{\eps}{1+\frac{1-\Omega}{K}V_2-\beps\Omega}
\right].\nonumber\\
\label{eq:shift15}\end{eqnarray}
Note that we have not yet taken the $\eps\to 0$ limit, so this formula is 
valid for all $\eps$. Now we simplify further by taking $\eps\to 0$, and a 
short calculation shows that the first three terms vanish. 
Therefore consider the contribution from the fourth term: 
\begin{equation}\beps \Tr_\mathcal{V}\left[
\frac{\eps}{1+V_1\frac{1-\Omega}{K}-\beps\Omega}\frac{K\Omega}{1-\Omega}
B c\Omega c\right]=\eps\beps \sum_{L=0}^\infty \beps^L
\Tr_\mathcal{V}\left[\left(\Omega-\frac{1}{\beps}V_1\frac{1-\Omega}{K}\right)^L
\frac{K\Omega}{1-\Omega}Bc\Omega c\right].\label{eq:shift2}
\end{equation}
Expanding the summand perturbatively in $V_1$,
\begin{eqnarray}\lineup\!\!\!\!\!\!\!\! \Tr_\mathcal{V}\left[\left(\Omega-\frac{1}{\beps}
V_1\frac{1-\Omega}{K}\right)^L
\frac{K\Omega}{1-\Omega}Bc\Omega c\right]\nonumber\\
\lineup=\Tr_\mathcal{V}\left[\Omega^L
\frac{K\Omega}{1-\Omega}Bc\Omega c\right]\nonumber\\
\lineup \ \ \ -\Tr_\mathcal{V}\left[\left(\sum_{k=0}^{L-1}\Omega^{L-1-k}
\left[\frac{1}{\beps}V_1
\frac{1-\Omega}{K}\right]\Omega^k\right)
\frac{K\Omega}{1-\Omega}Bc\Omega c\right]\nonumber\\
\lineup\ \ \ 
+\Tr_\mathcal{V}\left[\left(\sum_{k=0}^{L-2}\sum_{l=0}^{L-2-k}\Omega^{L-2-k-l}
\left[\frac{1}{\beps}V_1\frac{1-\Omega}{K}\right]\Omega^k
\left[\frac{1}{\beps}V_1
\frac{1-\Omega}{K}\right]\Omega^l\right)
\frac{K\Omega}{1-\Omega}Bc\Omega c\right]\nonumber\\
\lineup \ \ \ -\ .\ .\ .\ .
\end{eqnarray}
To derive the $\eps\to 0$ limit we expand this expression around 
$L=\infty$. Then each term in the perturbative expansion is a correlator
on a very large cylinder, and we can pick out the leading $\mathcal{L}^-$ level
of every field in the trace whose total width is fixed in the 
$L\to\infty$ limit. This gives:
\begin{eqnarray}
\lineup \!\!\!\!\!\!\!\Tr_\mathcal{V}\left[\left(\Omega-\frac{1}{\beps}
V_1\frac{1-\Omega}{K}\right)^L
\frac{K\Omega}{1-\Omega}Bc\Omega c\right]\nonumber\\
\lineup\  =-\Tr_\mathcal{V}[\Omega^LBc\d c]+\frac{1}{\beps}
\Tr_\mathcal{V}\left[\sum_{k=0}^{L-1}\Omega^{L-1-k}\,V_1\,\Omega^k\,
Bc\d c\right]\nonumber\\
\lineup\ \ \ -\frac{1}{\beps^2}\Tr_\mathcal{V}\left[
\sum_{k=0}^{L-2}\sum_{l=0}^{L-2-k}\Omega^{L-2-k-l}\,
V_1\, \Omega^k\, V_1\, \Omega^l\,
Bc\d c\right] +...+\mathcal{O}\left(\frac{1}{L}\right).
\end{eqnarray}
Scaling the circumference of the cylinders with $\frac{1}{L}$, the sums above
turn into integrals which precisely reproduce the boundary interaction of 
the marginal current \cite{KOS}:
\begin{eqnarray}
\lineup
\Tr_\mathcal{V}\left[\left(\Omega-\frac{1}{\beps}
V_1\frac{1-\Omega}{K}\right)^L
\frac{K\Omega}{1-\Omega}Bc\Omega c\right]\nonumber\\
\lineup\ \ =-\Tr_\mathcal{V}[\Omega
Bc\d c]+\frac{1}{\beps}\int_0^1 dx
\Tr_\mathcal{V}[\Omega^{1-x}\,V_1\,\Omega^x\,
Bc\d c]\nonumber\\
\lineup\ \ \ \ \ 
-\frac{1}{\beps^2}\int_0^1 dx\int_0^{1-y}dy\, 
\Tr_\mathcal{V}[\Omega^{1-x-y}\,
V_1\, \Omega^x\, V_1\, \Omega^y\, Bc\d c]+...
+\mathcal{O}\left(\frac{1}{L}\right)\nonumber\\
\lineup\ \ = -\Tr_\mathcal{V}[e^{-(K+\frac{1}{\beps}V_1)}Bc\d c]
+\mathcal{O}\left(\frac{1}{L}\right).
\end{eqnarray}
Plugging this back into \eq{shift2} and taking the $\eps\to0$ limit gives
\begin{equation}\lim_{\eps\to 0}\beps \Tr_\mathcal{V}\left[
\frac{\eps}{1+V_1\frac{1-\Omega}{K}-\beps\Omega}\frac{K\Omega}{1-\Omega}
B c\Omega c\right]=-\Tr_\mathcal{V}[e^{-(K+V_1)}Bc\d c].
\end{equation}
A similar argument for the final term in \eq{shift15}, and removing the 
$B$ ghost following \eq{Bghost} gives
\begin{eqnarray}\lim_{\eps\to 0}\Tr_\mathcal{V}[\psi_{12}(\eps)] \lineup = 
-\Tr_\mathcal{V}[e^{-(K+V_2)}c]+\Tr_\mathcal{V}[e^{-(K+V_1)}c]\nonumber\\
\lineup = \mathcal{A}_2(\mathcal{V})-\mathcal{A}_1(\mathcal{V}).
\end{eqnarray}
which is the expected shift in the closed string tadpole amplitude between 
the two marginal solutions.

\bigskip

\noindent {\bf Acknowledgments}

\bigskip

\noindent T.E. gratefully acknowledges travel support from a joint 
JSPS-MSMT grant LH11106. This research was funded by the EURYI grant 
GACR EYI/07/E010 from EUROHORC and ESF.


\begin{thebibliography}{10}

\bibitem{Schnabl}
  M.~Schnabl,
  ``Analytic solution for tachyon condensation in open string field 
	theory,''
  Adv.\ Theor.\ Math.\ Phys.\  {\bf 10}, 433 (2006)
  [arXiv:hep-th/0511286].

\bibitem{Okawa}
  Y.~Okawa,
  ``Comments on Schnabl's analytic solution for tachyon condensation in
  Witten's open string field theory,''
  JHEP {\bf 0604}, 055 (2006)
  [arXiv:hep-th/0603159].

\bibitem{FKsol}
  E.~Fuchs and M.~Kroyter,
  ``On the validity of the solution of string field theory,''
  JHEP {\bf 0605}, 006 (2006)
  [arXiv:hep-th/0603195].

\bibitem{SSF2}
  T.~Erler,
  ``Split string formalism and the closed string vacuum. II,''
  JHEP {\bf 0705}, 084 (2007)
  arXiv:hep-th/0612050.

\bibitem{Erler}
  T.~Erler,
  ``Tachyon Vacuum in Cubic Superstring Field Theory,''
  JHEP {\bf 0801}, 013 (2008)
  [arXiv:0707.4591 [hep-th]].

\bibitem{Russians1}
  I.~Y.~.Aref'eva, R.~V.~Gorbachev, D.~A.~Grigoryev, P.~N.~Khromov, 
  M.~V.~Maltsev and P.~B.~Medvedev,
  ``Pure Gauge Configurations and Tachyon Solutions to String Field 
  Theories Equations of Motion,''
  JHEP {\bf 0905}, 050 (2009)
  [arXiv:0901.4533 [hep-th]].

\bibitem{Russians2} 
  I.~Y.~.Aref'eva, R.~V.~Gorbachev and P.~B.~Medvedev,
  ``Pure Gauge Configurations and Solutions to Fermionic Superstring Field 
   Theories Equations of Motion,''
  J.\ Phys.\ A A {\bf 42}, 304001 (2009)
  [arXiv:0903.1273 [hep-th]].

\bibitem{simple}
  T.~Erler and M.~Schnabl,
  ``A Simple Analytic Solution for Tachyon Condensation,''
  JHEP {\bf 0910}, 066 (2009)
  [arXiv:0906.0979 [hep-th]].

\bibitem{exotic}
  T.~Erler,
  ``Exotic Universal Solutions in Cubic Superstring Field Theory,''
  JHEP {\bf 1104}, 107 (2011)
  [arXiv:1009.1865 [hep-th]].

\bibitem{Integra} 
  T.~Erler and C.~Maccaferri,
  ``Connecting Solutions in Open String Field Theory with Singular Gauge Transformations,''
  arXiv:1201.5119 [hep-th].

\bibitem{Ellwood}
  I.~Ellwood,
  ``The Closed string tadpole in open string field theory,''
  JHEP {\bf 0808}, 063 (2008).
  [arXiv:0804.1131 [hep-th]].

\bibitem{rolling}
  I.~Ellwood,
  ``Rolling to the tachyon vacuum in string field theory,''
  JHEP {\bf 0712}, 028 (2007)
  [arXiv:0705.0013 [hep-th]].

\bibitem{RZOK}
  M.~Kiermaier, Y.~Okawa, L.~Rastelli and B.~Zwiebach,
  ``Analytic solutions for marginal deformations in open string field 
  theory,''
  JHEP {\bf 0801}, 028 (2008)
  [arXiv:hep-th/0701249].

\bibitem{Schnabl2}
  M.~Schnabl,
  ``Comments on marginal deformations in open string field theory,''
  Phys.\ Lett.\  B {\bf 654}, 194 (2007)
  [arXiv:hep-th/0701248].

\bibitem{Kishimoto} 
  I.~Kishimoto,
  ``Comments on gauge invariant overlaps for marginal solutions in open 
  string field theory,''
  Prog.\ Theor.\ Phys.\  {\bf 120}, 875 (2008)
  [arXiv:0808.0355 [hep-th]].

\bibitem{RSZproj}
  L.~Rastelli, A.~Sen and B.~Zwiebach,
  ``Boundary CFT construction of D-branes in vacuum string field theory,''
  JHEP {\bf 0111}, 045 (2001)
  [hep-th/0105168].

\bibitem{boundary}
  M.~Kiermaier, Y.~Okawa and B.~Zwiebach,
  ``The boundary state from open string fields,''
  [arXiv:0810.1737 [hep-th]].

\bibitem{SSF1}
  T.~Erler,
  ``Split string formalism and the closed string vacuum,''
  JHEP {\bf 0705}, 083 (2007)
  [arXiv:hep-th/0611200].

\bibitem{RZ}
  L.~Rastelli and B.~Zwiebach,
  ``Solving open string field theory with special projectors,''
  JHEP {\bf 0801}, 020 (2008)
  arXiv:hep-th/0606131.

\bibitem{id1}
  E.~A.~Arroyo,
  ``Generating Erler-Schnabl-type Solution for Tachyon Vacuum in Cubic
  Superstring Field Theory,''
  J.\ Phys.\ A  {\bf 43}, 445403 (2010)
  [arXiv:1004.3030 [hep-th]].

\bibitem{id2} 
  S.~Zeze,
  ``Regularization of identity based solution in string field theory,''
  JHEP {\bf 1010}, 070 (2010)
  [arXiv:1008.1104 [hep-th]].

\bibitem{bcc}
  M.~Kiermaier, Y.~Okawa and P.~Soler,
  ``Solutions from boundary condition changing operators in open 
  string field theory,''
  JHEP {\bf 1103}, 122 (2011)
  [arXiv:1009.6185 [hep-th]].

\bibitem{BMT}
  L.~Bonora, C.~Maccaferri and D.~D.~Tolla,
  ``Relevant Deformations in Open String Field Theory: a Simple Solution 
  for Lumps,''
  JHEP\ {\bf 1111}, 107  (2011)
  [arXiv:1009.4158 [hep-th]].

\bibitem{KOS}
  M.~Kiermaier, Y.~Okawa and P.~Soler,
  ``Solutions from boundary condition changing operators in open 
  string field theory,''
  JHEP {\bf 1103}, 122 (2011)
  [arXiv:1009.6185 [hep-th]].


\bibitem{KOS_super}
  T.~Noumi and Y.~Okawa,
  ``Solutions from boundary condition changing operators in open superstring 
  field theory,''
  JHEP {\bf 1112}, 034 (2011)
  [arXiv:1108.5317 [hep-th]].


\end{thebibliography}
\end{document}